\documentclass[preprint,12pt]{elsarticle}

\usepackage{amssymb}

\journal{Physics Letters A}

\begin{document}

\begin{frontmatter}



\title{Two-particle Irreducible Effective Action Approach to Correlated Electron Systems}


\author[label1,label2]{Wei-Jie Fu}
\ead{fuw@brandonu.ca}

\address[label1]{Department of Physics,
Brandon University, Brandon, Manitoba, R7A 6A9 Canada}
\address[label2]{Winnipeg Institute for Theoretical Physics, Winnipeg,
Manitoba}

\begin{abstract}
The two-particle irreducible (2PI) effective action theories are
employed to study the strongly fluctuating electron systems, under
the formalism of the two-dimensional Hubbard model. We obtain the
corresponding quantum 2PI effective action after the original
classic action of the Hubbard model is bosonized. In our actual
calculations, the 2PI effective action is expanded to three loops,
in which the leading order (LO) and next-to-leading order (NLO)
quantum fluctuations are included. Numerical calculations indicate
that the NLO fluctuations should not be neglected when the Coulomb
on-site repulsion energy is larger than two times the
nearest-neighbor hopping energy.
\end{abstract}


\end{frontmatter}


\section{Introduction}
\vspace{5pt}

Since the high-temperature superconductivity was discovered in
1986~\cite{Bednorz1986}, It has been believed that the appropriate
model to describe the strongly correlated electron systems is the
nearly half-filled two-dimensional Hubbard model with moderately
large repulsion energy $U$ and antiferromagnetic exchange constant
$J=4t^{2}/U$ where $t$ is the site hopping~\cite{Anderson1987}. The
Hubbard model is just composed of two terms: one is the site-hopping
term which forms the band structure, and the other is the Coulomb
repulsion term that represents the interaction between electrons.
Therefore, from its appearance, it looks like that the
two-dimensional Hubbard model is an simple model and it is easy to
solved. However, the actual situations are quite different. The
Hubbard model describes a many-body electron system in which the
interacting potential energy and the kinetic energy are comparable.
So we can not employ the perturbation theory and treat the potential
energy or the kinetic energy as a perturbation.

Lots of efforts have been made to solve the two-dimensional Hubbard
model and many methods have been developed. For example, the quantum
Monte Carlo simulations~\cite{Hirsch1988}, the self-consistent
approach of conserving approximations~\cite{Bickers1989}, the
variational cluster perturbation theory~\cite{Tremblay2005}, the
functional renormalization group approach~\cite{Honerkamp2007}, and
so on. In recent years, another nonperturbative approach, known as
the two-particle irreducible (2PI) effective action theory first
introduced in the field theory~\cite{Cornwall1974}, has attracted
lots of attentions. The 2PI effective action theory resums certain
classes of diagrams to infinite order, so nonperturbative effects
are included in this approach. Furthermore, in the 2PI formalism,
the effective action can be expanded according to the order of the
loop or $1/N$ in the $O(N)$ model. Therefore, it is easy to
investigate the effects of the high order contributions in the 2PI
effective action theory. In the studies of field theories, it has
been found that the 2PI effective action theory is very successful
in describing equilibrium thermodynamics, and also the quantum
dynamics of far from equilibrium of quantum fields. The entropy of
the quark-gluon plasma obtained in the 2PI formalism shows very good
agreement with lattice data for temperatures above twice the
transition temperature~\cite{Blaizot1999}. The poor convergence
problem usually encountered in high-temperature resummed
perturbation theory with bosonic fields is also solved in the 2PI
effective action theory~\cite{Berges2005a}. Furthermore, it has been
shown that non-equilibrium dynamics with subsequent late-time
thermalization can be well described in the 2PI formalism (see
\cite{Berges2001} and references therein). The 2PI effective action
has also been combined with the exact renormalization group to
provide efficient non-perturbative approximation
schemes~\cite{Blaizot2011}. The shear viscosity in the $O(N)$ model
has been computed using the 2PI formalism~\cite{Aarts2004}.
Specially, we would like to emphasize that due to many people's
contributions~\cite{van2002,Blaizot2004,Berges2005,Reinosa2010}, it
has been clear that the 2PI effective action theory can be
renormalized, which is quite non-trivial for a non-perturbative
approach.

In this work, we will employ the 2PI effective action theory to
investigate the strong fluctuations of electron systems. We will
expand the effective action to three loops and compute the leading
order (LO) and the next-to-leading order (NLO) contributions to the
fermion and boson self-energies. Then we will investigate when the
importance of the high order quantum fluctuations becomes
significant with the increase of the Coulomb repulsion energy $U$.
The paper is organized as follows. In section \ref{2PIsection} we
apply the 2PI formalism into the Hubbard model and obtain its
effective action. In section \ref{SelfEnergysection} we obtain the
LO and NLO contributions to the fermion and boson self-energies.
Numerical results are presented in Sec. \ref{Numericalsection}. In
section \ref{concSect} we give our summary and conclusions.

\section{2PI Effective Action Theory}
\label{2PIsection} \vspace{5pt}

We begin with the simplest two-dimensional one-band Hubbard model
which reads
\begin{equation}
H=-t\sum_{\langle
ij\rangle\sigma}{\hat{c}}^{\dag}_{i\sigma}{\hat{c}}_{j\sigma}
+U\sum_{i}{\hat{c}}^{\dag}_{i\uparrow}{\hat{c}}_{i\uparrow}{\hat{c}}^{\dag}_{i\downarrow}{\hat{c}}_{i\downarrow},
\end{equation}
where we only consider the nearest-neighbor hopping $t$ and $U$ is
the Hubbard on-site Coulomb repulsion energy. We rewrite the
interaction term as
\begin{equation}
{\hat{c}}^{\dag}_{i\uparrow}{\hat{c}}_{i\uparrow}{\hat{c}}^{\dag}_{i\downarrow}{\hat{c}}_{i\downarrow}
=-\frac{1}{2}({\hat{c}}^{\dag}_{i}\sigma^{z}{\hat{c}}_{i})^{2}+\frac{1}{2}({\hat{c}}^{\dag}_{i\uparrow}{\hat{c}}_{i\uparrow}
+{\hat{c}}^{\dag}_{i\downarrow}{\hat{c}}_{i\downarrow})\label{cccc}
\end{equation}
for the convenience of calculations below, where $\sigma^{z}$ is the
$z$ component of the Pauli matrices. The second term on the right
hand side of Eq.~(\ref{cccc}) can be absorbed in the chemical
potential term. Then we arrive at
\begin{equation}
H-\mu N=-t\sum_{\langle
ij\rangle\sigma}{\hat{c}}^{\dag}_{i\sigma}{\hat{c}}_{j\sigma}
-\frac{U}{2}\sum_{i}({\hat{c}}^{\dag}_{i}\sigma^{z}{\hat{c}}_{i})^{2}
-\mu\sum_{
i\sigma}{\hat{c}}^{\dag}_{i\sigma}{\hat{c}}_{i\sigma}.\label{Hamiltonian}
\end{equation}
The classic action corresponding to Hamiltonian given above is
\begin{equation}
S=\int dt\bigg[\sum_{
i\sigma}c^{*}_{i\sigma}i\partial_{t}c_{i\sigma}+t\sum_{\langle
ij\rangle\sigma}c^{*}_{i\sigma}c_{j\sigma} +\mu\sum_{
i\sigma}c^{*}_{i\sigma}c_{i\sigma}+\frac{U}{2}\sum_{i}(c^{\dag}_{i}\sigma^{z}c_{i})^{2}\bigg]
,\label{}
\end{equation}
where creating and annihilating operators in Eq.~(\ref{Hamiltonian})
are replaced by their Grassmann fields. Including quantum and
thermal fluctuations, one obtains the generating functional, also
known as the partition function, which reads
\begin{eqnarray}
Z[\eta^{*},\eta,J]&=&\int
[dc^{*}][dc]\exp\Bigg\{-\int_{0}^{\beta}d\tau \bigg[\sum_{
i\sigma}c^{*}_{i\sigma}\partial_{\tau}c_{i\sigma}-t\sum_{\langle
ij\rangle\sigma}c^{*}_{i\sigma}c_{j\sigma}\nonumber \\&& -\mu\sum_{
i\sigma}c^{*}_{i\sigma}c_{i\sigma}
-\frac{U}{2}\sum_{i}(c^{\dag}_{i}\sigma^{z}c_{i})^{2} +\sum_{
i\sigma}(\eta_{i\sigma}^{*}c_{i\sigma}+c_{i\sigma}^{*}\eta_{i\sigma})
\nonumber
\\&&+\sum_{i}J_{i}(-1)^{i}c^{\dag}_{i}\sigma^{z}c_{i}\bigg]\Bigg\}.\label{GeneFunc}
\end{eqnarray}
We will employ the Matsubara imaginary-time formalism throughout
this work and here $\beta=1/T$ is the inverse of the temperature. An
external source term for the composite operator
$(-1)^{i}c^{\dag}_{i}\sigma^{z}c_{i}$ is included in
Eq.~(\ref{GeneFunc}).

Before we continue the calculations, it would be more convenient if
the bosonization of the action in Eq.~(\ref{GeneFunc}) is made
first. Up to a constant, we have
\begin{eqnarray}
&&\exp\Bigg\{-\int_{0}^{\beta}d\tau\bigg[-\frac{U}{2}\sum_{i}(c^{\dag}_{i}\sigma^{z}c_{i})^{2}
+\sum_{i}J_{i}(-1)^{i}c^{\dag}_{i}\sigma^{z}c_{i}\bigg]\Bigg\}\nonumber
\\&=&\exp\bigg[-\frac{1}{2U}\int_{0}^{\beta}d\tau\sum_{i}J_{i}^{2}\bigg]
\int[dB]\exp\Bigg\{-\int_{0}^{\beta}d\tau\sum_{i}\bigg[\frac{U}{2}B_{i}^{2}\nonumber
\\&& -UB_{i}(-1)^{i}c^{\dag}_{i}\sigma^{z}c_{i}+J_{i}B_{i}\bigg]\Bigg\},\label{FuncE}
\end{eqnarray}
where a boson field $B$ is introduced through its functional
integral. Substituting Eq.~(\ref{FuncE}) into Eq.~(\ref{GeneFunc})
and neglecting the irrelevant prefactor on the right hand side of
Eq.~(\ref{FuncE}), one finds
\begin{eqnarray}
Z[\eta^{*},\eta,J]&=&\int
[dc^{*}][dc][dB]\exp\Bigg\{-\int_{0}^{\beta}d\tau \bigg[\sum_{
i\sigma}c^{*}_{i\sigma}\partial_{\tau}c_{i\sigma}-t\sum_{\langle
ij\rangle\sigma}c^{*}_{i\sigma}c_{j\sigma} \nonumber \\&&-\mu\sum_{
i\sigma}c^{*}_{i\sigma}c_{i\sigma}
+\sum_{i}\Big(\frac{U}{2}B_{i}^{2}-UB_{i}(-1)^{i}c^{\dag}_{i}\sigma^{z}c_{i}\Big)
\nonumber \\&&+\sum_{
i\sigma}(\eta_{i\sigma}^{*}c_{i\sigma}+c_{i\sigma}^{*}\eta_{i\sigma})
+\sum_{i}J_{i}B_{i}\bigg]\Bigg\}.\label{GeneFunc2}
\end{eqnarray}

In order to obtain the 2PI effective action for the Hubbard model,
we need add two-point sources into Eq.~(\ref{GeneFunc2}). Then we
obtain
\begin{eqnarray}
Z[\eta^{*},\eta,J,M,K]&=&\int
[dc^{*}][dc][dB]\exp\bigg\{-[I_{0}(c^{*},c)+I_{0}(B)+I_{\mathbf{int}}(c^{*},c,B)\nonumber
\\&&+\eta^{*}c+c^{*}\eta+JB+\frac{1}{2}BMB+c^{*}Kc]\bigg\},\label{GeneFunc3}
\end{eqnarray}
where the two-point external sources are
\begin{eqnarray}
\frac{1}{2}BMB&\equiv&\frac{1}{2}\int_{0}^{\beta}d\tau_{i}d\tau_{j}\sum_{ij}B_{i}(\tau_{i})
M_{ij}(\tau_{i},\tau_{j})B_{j}(\tau_{j}),\\
c^{*}Kc&\equiv&\int_{0}^{\beta}d\tau_{i}d\tau_{j}\sum_{ij}\sum_{\alpha\beta}c_{i\alpha}^{*}(\tau_{i})
K_{i\alpha,j\beta}(\tau_{i},\tau_{j})c_{j\beta}(\tau_{j}).\label{}
\end{eqnarray}
Here $\alpha$ and $\beta$ are spin indices. In Eq.~(\ref{GeneFunc3})
we also used the following abbreviated notations:
\begin{eqnarray}
I_{0}(c^{*},c)&\equiv&\int_{0}^{\beta}d\tau\bigg[\sum_{
i\alpha}c^{*}_{i\alpha}\partial_{\tau}c_{i\alpha}-t\sum_{\langle
ij\rangle\alpha}c^{*}_{i\alpha}c_{j\alpha} -\mu\sum_{
i\alpha}c^{*}_{i\alpha}c_{i\alpha}\bigg],\\
I_{0}(B)&\equiv&\int_{0}^{\beta}d\tau\sum_{
i}\frac{U}{2}B_{i}^{2},\\
I_{\mathrm{int}}(c^{*},c,B)&\equiv&\int_{0}^{\beta}d\tau\sum_{
i}(-U)B_{i}(-1)^{i}c^{\dag}_{i}\sigma^{z}c_{i},\\
\eta^{*}c+c^{*}\eta&\equiv&\int_{0}^{\beta}d\tau\sum_{i\alpha}
(\eta_{i\alpha}^{*}c_{i\alpha}+c_{i\alpha}^{*}\eta_{i\alpha}),\\
JB&\equiv&\int_{0}^{\beta}d\tau\sum_{i}J_{i}B_{i}.\label{}
\end{eqnarray}
Introducing the generating functional for the connected Green
functions
\begin{equation}
W[\eta^{*},\eta,J,M,K]=-\ln Z[\eta^{*},\eta,J,M,K],\label{}
\end{equation}
it then follows that
\begin{eqnarray}
\frac{\delta W}{\delta J_{i}}&=&B_{i}^{c},\quad \frac{\delta
W}{\delta \eta_{i\alpha}^{*}}=c_{i\alpha}^{c},\quad \frac{\delta
W}{\delta \eta_{i\alpha}}=-c_{i\alpha}^{*c},
\\\frac{\delta W}{\delta M_{ji}}&=&\frac{1}{2}(B_{i}^{c}B_{j}^{c}+G_{ij}),
\quad \frac{\delta W}{\delta
K_{j\beta,i\alpha}}=-(c_{i\alpha}^{c}c_{j\beta}^{*c}+S_{i\alpha,j\beta}),\label{}
\end{eqnarray}
where $B^{c}$, $c^{c}$, and $c^{*c}$ are the expected values of
fields $B$, $c$, and $c^{*}$, respectively. $G$ and $S$ are the
propagators for boson and fermion fields.

The 2PI effective action can be obtained from $W$ through the
Legendre transformation as follows
\begin{eqnarray}
\Gamma[c^{c},c^{*c},B^{c},G,S]&=&W[\eta^{*},\eta,J,M,K]-J_{i}\frac{\delta
W}{\delta J_{i}}-\eta_{i\alpha}^{*}\frac{\delta W}{\delta
\eta_{i\alpha}^{*}}-\eta_{i\alpha}\frac{\delta W}{\delta
\eta_{i\alpha}}\nonumber
\\&&-M_{ji}\frac{\delta W}{\delta
M_{ji}}-K_{j\beta,i\alpha}\frac{\delta W}{\delta
K_{j\beta,i\alpha}}\nonumber
\\&=&W[\eta^{*},\eta,J,M,K]-J_{i}B_{i}^{c}-\eta_{i\alpha}^{*}c_{i\alpha}^{c}-c_{i\alpha}^{*c}\eta_{i\alpha}\nonumber
\\&&-\frac{1}{2}\mathrm{Tr}[M(B^{c}B^{c}+G)]+\mathrm{Tr}[K(c^{c}c^{*c}+S)],\label{}
\end{eqnarray}
where summations and integrals are assumed for the repeated indices.
The trace operates in the coordinate and inner spaces. It can be
easily proved that
\begin{eqnarray}
\frac{\delta \Gamma}{\delta
B_{i}^{c}}&=&-J_{i}-M_{ij}B_{j}^{c},\quad \frac{\delta
\Gamma}{\delta
c_{i\alpha}^{*c}}=-\eta_{i\alpha}-K_{i\alpha,j\beta}c^{c}_{j\beta},\label{SelfEq1}\\
\frac{\delta\Gamma}{\delta
c_{i\alpha}^{c}}&=&\eta_{i\alpha}^{*}+c^{*c}_{j\beta}K_{j\beta,i\alpha},\quad
\frac{\delta \Gamma}{\delta G_{ij}}=-\frac{1}{2}M_{ji},\quad
\frac{\delta \Gamma}{\delta
S_{i\alpha,j\beta}}=K_{j\beta,i\alpha}.\label{SelfEq2}
\end{eqnarray}
Equations~(\ref{SelfEq1}) and (\ref{SelfEq2}) form a set of
self-consistent equations which determine the field expected values
$B^{c}$, $c^{c}$, $c^{*c}$ and the propagators $G$ and $S$, if the
effective action can be expressed as a functional of these field
expected values and propagators. Usually, the expected values of
fermion field $c^{c}$ and $c^{*c}$ are vanishing when their external
sources $\eta^{*}$ and $\eta$ are zero. We will assume
$c^{c}=c^{*c}=0$ in the following calculations and use $B$ in place
of $B^{c}$ without confusions. It can be shown that the 2PI
effective action can be expressed as~\cite{Cornwall1974}
\begin{eqnarray}
\Gamma(B,G,S)&=&I(B)+\frac{1}{2}\mathrm{Tr}\ln
G^{-1}+\frac{1}{2}\mathrm{Tr}(G_{0}^{-1}G)\nonumber
\\&&-\mathrm{Tr}\ln S^{-1}-\mathrm{Tr}(S_{0}^{-1}S)+\Gamma_{\mathrm{int}}(B,G,S),\label{Gamma}
\end{eqnarray}
with
\begin{equation}
\Gamma_{\mathrm{int}}(B,G,S)=-\mathrm{Tr}[(-\Sigma_{\mathrm{mean}})S]+\Gamma_{2}(G,S).\label{}
\end{equation}
Here we have $I(B)=I_{0}(B)$ and
\begin{eqnarray}
G_{0}^{-1}\!\!\!\!&=&\!\!\!\!\frac{\delta^{2}I_{0}(B)}{\delta
B^{2}},\quad S_{0}^{-1}=-\frac{\delta^{2}I_{0}(c^{*},c)}{\delta
c^{*}\delta c},\quad
-\Sigma_{\mathrm{mean}}=-\frac{\delta^{2}I_{\mathrm{int}}(c^{*},c,B)}{\delta
c^{*}\delta c},\label{G0S0}\\
\Gamma_{2}(G,S)\!\!\!\!\!&=&\!\!\!\!\!-\ln\!\!\frac{\int
[dc^{*}][dc][dB]\exp\bigg\{\!\!-\!\!\Big[\frac{1}{2}BG^{-1}B\!\!+\!\!c^{*}S^{-1}c\!+\!\!I_{\mathrm{int}}(c^{*},c,B)\Big]\bigg\}}
{\int
[dc^{*}][dc][dB]\exp\bigg\{-\Big[\frac{1}{2}BG^{-1}B+c^{*}S^{-1}c\Big]\bigg\}}\Bigg|_{\mathrm{2PI}},\label{Gamma2}
\end{eqnarray}
where $\Sigma_{\mathrm{mean}}$ is the mean field contribution to the
fermion self-energy; $\Gamma_{2}$ sums all 2PI diagrams. The
prominent difference between these diagrams and the perturbative
ones is that the propagators constituting these diagrams are
self-consistent ones $G$ and $S$, not $G_{0}$ and $S_{0}$. But the
vertices are bare, which are same as those in perturbative diagrams.

\section{Self-Energy of Fermion and Boson Fields}
\label{SelfEnergysection}

\begin{figure}[!htb]
\includegraphics[scale=0.6]{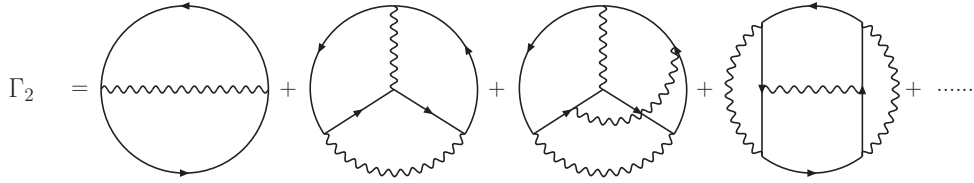}
\caption{Two-particle irreducible vacuum diagrams contributing to
the effective action, where the solid and wavy lines represent
fermion and boson propagators, respectively.}\label{Gamma2Fig}
\end{figure}

$\Gamma_{2}$ in Eq.~(\ref{Gamma2}) receives contributions from
infinite 2PI diagrams, some of which are shown in
Fig.~\ref{Gamma2Fig}. In actual calculations, it is impossible to
sum all these diagrams. We have to truncate the effective action to
certain order (But it is possible to resum some kinds of diagrams to
infinite order through other methods, see Ref.~\cite{Carrington2012}
for more details). In this work we will truncate the effective
action to three loops, i.e., only the first two diagram in
Fig.~\ref{Gamma2Fig} are employed in our calculations.

\begin{figure}[!htb]
\includegraphics[scale=0.6]{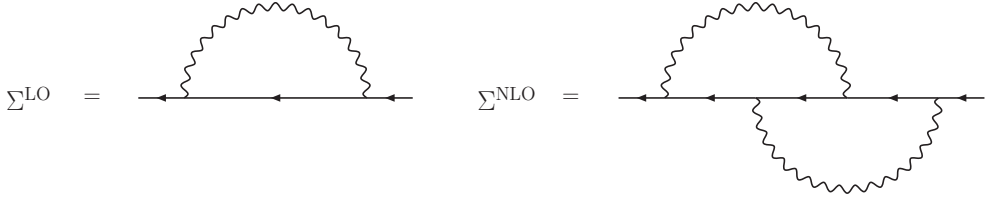}
\caption{LO and NLO contributions to the fermion
self-energy.}\label{SigmaFig}
\end{figure}

\begin{figure}[!htb]
\includegraphics[scale=0.6]{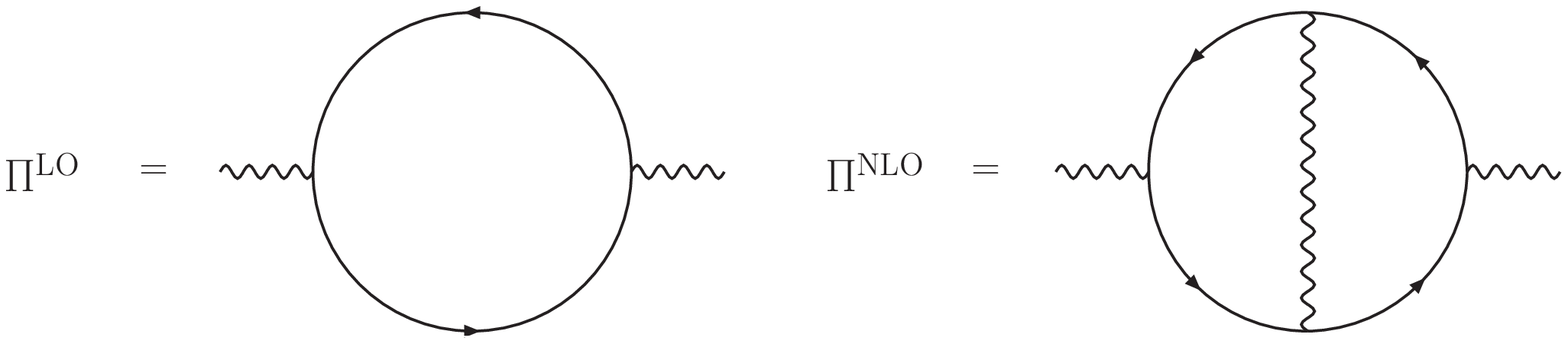}
\caption{LO and NLO contributions to the boson
self-energy.}\label{PiFig}
\end{figure}

In the following we will call the two-loop and three-loop diagrams
as the LO and NLO contributions to $\Gamma_{2}$, respectively. Their
expressions are

\begin{eqnarray}
\Gamma_{2}^{\mathrm{LO}}&=&\int_{0}^{\beta}d\tau_{1}d\tau_{2}\sum_{i_{1}i_{2}}
(-1)^{i_{1}+i_{2}}\frac{1}{2}U^{2}G_{\tau_{1}i_{1},\tau_{2}i_{2}}
\mathrm{tr}(\sigma^{z}S_{\tau_{2}i_{2},\tau_{1}i_{1}}
\sigma^{z}S_{\tau_{1}i_{1},\tau_{2}i_{2}}),
\\\Gamma_{2}^{\mathrm{NLO}}&=&\int_{0}^{\beta}d\tau_{1}d\tau_{2}d\tau_{3}d\tau_{4}\sum_{i_{1}i_{2}i_{3}i_{4}}
(-1)^{i_{1}+i_{2}+i_{3}+i_{4}}\frac{1}{4}U^{4}G_{\tau_{1}i_{1},\tau_{3}i_{3}}
G_{\tau_{2}i_{2},\tau_{4}i_{4}}\nonumber
\\&&\times\mathrm{tr}(\sigma^{z}S_{\tau_{4}i_{4},\tau_{1}i_{1}}\sigma^{z}S_{\tau_{1}i_{1},\tau_{2}i_{2}}
\sigma^{z}S_{\tau_{2}i_{2},\tau_{3}i_{3}}\sigma^{z}S_{\tau_{3}i_{3},\tau_{4}i_{4}}),\label{}
\end{eqnarray}
where the trace $\mathrm{tr}$ only operates in spin space. Up to
now, we have expressed the effective action as a functional of the
self-consistent propagators $G$ and $S$. Then one can employ
Eq.~(\ref{SelfEq2}) to obtain the self-consistent equations, given
by
\begin{eqnarray}
S^{-1}&=&S^{-1}_{0}-\Sigma,\label{FermiEq}
\\G^{-1}&=&G^{-1}_{0}-\Pi,\label{BoseEq}
\end{eqnarray}
where the fermion and boson self-energies are
\begin{eqnarray}
\Sigma&=&\frac{\delta \Gamma_{\mathrm{int}}}{\delta
S}=\Sigma_{\mathrm{mean}}+\frac{\delta \Gamma_{2}(G,S)}{\delta
S}\nonumber
\\&\approx &\Sigma_{\mathrm{mean}}+\Sigma^{\mathrm{LO}}(G,S)+\Sigma^{\mathrm{NLO}}(G,S),
\\\Pi&=&-2\frac{\delta \Gamma_{\mathrm{int}}}{\delta
G}=-2\frac{\delta \Gamma_{2}(G,S)}{\delta G}\nonumber
\\&\approx &\Pi^{\mathrm{LO}}(G,S)+\Pi^{\mathrm{NLO}}(G,S).\label{}
\end{eqnarray}
They are depicted in Fig.~\ref{SigmaFig} and Fig.~\ref{PiFig},
respectively, whose expressions read
\begin{eqnarray}
\Sigma^{\mathrm{LO}}_{\tau_{1}i_{1}\alpha_{1},\tau_{2}i_{2}\alpha_{2}}&=&\sum_{\beta_{1}\beta_{2}}(-1)^{i_{1}+i_{2}}
U^{2}G_{\tau_{1}i_{1},\tau_{2}i_{2}}\sigma^{\mathrm{z}}_{\alpha_{1}\beta_{1}}S_{\tau_{1}i_{1}\beta_{1},\tau_{2}i_{2}\beta_{2}}
\sigma^{\mathrm{z}}_{\beta_{2}\alpha_{2}},
\label{SEFLOC}\\\Sigma^{\mathrm{NLO}}_{\tau_{1}i_{1}\alpha_{1},\tau_{2}i_{2}\alpha_{2}}&=&\int_{0}^{\beta}d\tau_{3}d\tau_{4}
\sum_{i_{3}i_{4}}\sum_{\alpha_{3}\alpha_{4}}
\sum_{\beta_{1}...\beta_{4}}(-1)^{i_{1}+i_{2}+i_{3}+i_{4}}U^{4}
G_{\tau_{1}i_{1},\tau_{3}i_{3}}\nonumber
\\&&\times G_{\tau_{4}i_{4},\tau_{2}i_{2}}\sigma^{\mathrm{z}}_{\alpha_{1}\beta_{1}}
S_{\tau_{1}i_{1}\beta_{1},\tau_{4}i_{4}\alpha_{4}}\sigma^{\mathrm{z}}_{\alpha_{4}\beta_{4}}
S_{\tau_{4}i_{4}\beta_{4},\tau_{3}i_{3}\alpha_{3}}
\sigma^{\mathrm{z}}_{\alpha_{3}\beta_{3}}\nonumber
\\&&\times
S_{\tau_{3}i_{3}\beta_{3},\tau_{2}i_{2}\beta_{2}}\sigma^{\mathrm{z}}_{\beta_{2}\alpha_{2}},\\
\Pi^{\mathrm{LO}}_{\tau_{1}i_{1},\tau_{2}i_{2}}&=&-U^{2}(-1)^{i_{1}+i_{2}}\mathrm{tr}(\sigma^{z}S_{\tau_{1}i_{1},\tau_{2}i_{2}}
\sigma^{z}S_{\tau_{2}i_{2},\tau_{1}i_{1}}),\\
\Pi^{\mathrm{NLO}}_{\tau_{1}i_{1},\tau_{2}i_{2}}&=&\int_{0}^{\beta}d\tau_{3}d\tau_{4}
\sum_{i_{3}i_{4}}(-U^{4})G_{\tau_{3}i_{3},\tau_{4}i_{4}}
(-1)^{i_{1}+i_{2}+i_{3}+i_{4}}\nonumber
\\&&\times\mathrm{tr}(\sigma^{z}S_{\tau_{1}i_{1},\tau_{4}i_{4}}\sigma^{z}S_{\tau_{4}i_{4},\tau_{2}i_{2}}
\sigma^{z}S_{\tau_{2}i_{2},\tau_{3}i_{3}}\sigma^{z}S_{\tau_{3}i_{3},\tau_{1}i_{1}}).\label{SEBNLOC}
\end{eqnarray}

Up to now, we have worked in coordinate space. In fact, it is more
convenient to employ the momentum lattices, in which numerical
calculations are easier. The coordinate lattices and the momentum
ones are related through the Fourier transformation. For example,
for the boson and fermion propagators we have
\begin{eqnarray}
G_{\tau_{1}i_{1},\tau_{2}i_{2}}\!\!\!\!&=&\!\!\!\!\beta^{-2}\frac{1}{(\sqrt{N})^{2}}
\sum_{\omega_{1}k_{1}}\sum_{\omega_{2}k_{2}}G(\omega_{1}k_{1},\omega_{2}k_{2})
\exp(-i\omega_{1}\tau_{1}+i\omega_{2}\tau_{2}\nonumber
\\&&+ik_{1}R_{i_{1}}-ik_{2}R_{i_{2}}),
\label{FourB}\\S_{\tau_{1}i_{1}\alpha_{1},\tau_{2}i_{2}\alpha_{2}}\!\!\!\!&=&\!\!\!\!\beta^{-2}\frac{1}{(\sqrt{N})^{2}}
\sum_{\omega_{1}k_{1}}\sum_{\omega_{2}k_{2}}S_{\alpha_{1},\alpha_{2}}(\omega_{1}k_{1},\omega_{2}k_{2})\nonumber
\\&&\times
\exp(-i\omega_{1}\tau_{1}+i\omega_{2}\tau_{2}+ik_{1}R_{i_{1}}-ik_{2}R_{i_{2}}),\label{FourF}
\end{eqnarray}
where $N$ is the number of lattice sites, momentum sums are
restricted to the Brillouin zone, and the Matsubara frequencies are
\begin{equation}
\omega_{i}=\Bigg\{\begin{array}{ll} (2n_{i}+1)\pi T\quad &
\mathrm{(Fermion)}\\
2n_{i}\pi T & \mathrm{(Boson)}\end{array}.\label{}
\end{equation}

Employing Eqs.~(\ref{FourB}) and (\ref{FourF}), we can reexpress the
self-energies in Eqs.~(\ref{SEFLOC})---(\ref{SEBNLOC}) in momentum
lattices as follow
\begin{eqnarray}
\Sigma^{\mathrm{LO}}_{\alpha_{1},\alpha_{2}}(\bar{\omega}_{1}\bar{k}_{1},\bar{\omega}_{2}\bar{k}_{2})&=&\beta^{-2}\frac{1}{(\sqrt{N})^{2}}
\sum_{\omega_{1}k_{1}}\sum_{\omega_{2}k_{2}}\sum_{\alpha_{1}^{\prime}\alpha_{2}^{\prime}}U^{2}
G(\omega_{1}k_{1},\omega_{2}k_{2})
\sigma^{\mathrm{z}}_{\alpha_{1}\alpha_{1}^{\prime}}\nonumber
\\&&\times
S_{\alpha_{1}^{\prime},\alpha_{2}^{\prime}}(\bar{\omega}_{1}-\omega_{1},\bar{k}_{1}-k_{1}-Q;
\bar{\omega}_{2}-\omega_{2},\bar{k}_{2}-k_{2}-Q)\nonumber
\\&&\times
\sigma^{\mathrm{z}}_{\alpha_{2}^{\prime}\alpha_{2}}e^{-i(\bar{\omega}_{1}-\omega_{1})0^{+}},\label{SEFLOM}
\\\Sigma^{\mathrm{NLO}}_{\alpha_{1},\alpha_{2}}(\bar{\omega}_{1}\bar{k}_{1},\bar{\omega}_{2}\bar{k}_{2})&=&
\beta^{-6}\frac{1}{(\sqrt{N})^{4}}\sum_{\omega_{1}k_{1}...\omega_{4}k_{4}}\sum_{\omega_{3}^{\prime}k_{3}{\prime}}
\sum_{\omega_{4}^{\prime}k_{4}{\prime}}\sum_{\alpha_{3}\alpha_{4}}
\sum_{\alpha_{1}^{\prime}...\alpha_{4}^{\prime}}U^{4}\nonumber
\\&&\times G(\omega_{1}k_{1},\omega_{4}k_{4})G(\omega_{3}k_{3},\omega_{2}k_{2})
\nonumber
\\&&\times\sigma^{\mathrm{z}}_{\alpha_{1}\alpha_{1}^{\prime}}
S_{\alpha_{1}^{\prime},\alpha_{3}^{\prime}}(\bar{\omega}_{1}-\omega_{1},\bar{k}_{1}-k_{1}-Q;\omega_{3}^{\prime},k_{3}^{\prime})
\sigma^{\mathrm{z}}_{\alpha_{3}^{\prime}\alpha_{3}}\nonumber
\\&&\times
S_{\alpha_{3},\alpha_{4}}(\omega_{3}^{\prime}-\omega_{3},k_{3}^{\prime}-k_{3}-Q;\omega_{4}^{\prime}-\omega_{4},k_{4}^{\prime}-k_{4}-Q)
\nonumber
\\&&\times\sigma^{\mathrm{z}}_{\alpha_{4}\alpha_{4}^{\prime}}
S_{\alpha_{4}^{\prime},\alpha_{2}^{\prime}}(\omega_{4}^{\prime},k_{4}^{\prime};\bar{\omega}_{2}-\omega_{2},\bar{k}_{2}-k_{2}-Q)
\sigma^{\mathrm{z}}_{\alpha_{2}^{\prime}\alpha_{2}},\\
\Pi^{\mathrm{LO}}(\bar{\omega}_{1}\bar{k}_{1},\bar{\omega}_{2}\bar{k}_{2})&=&-\beta^{-2}\frac{1}{(\sqrt{N})^{2}}
\sum_{\omega_{1}k_{1}}\sum_{\omega_{2}k_{2}}U^{2}\mathrm{tr}\bigg[S(\omega_{1}k_{1},\omega_{2}k_{2})\sigma^{\mathrm{z}}
\nonumber
\\&&\times S(\omega_{2}\!\!-\!\!\bar{\omega}_{2},k_{2}-\bar{k}_{2}+Q;\omega_{1}\!\!-\!\!\bar{\omega}_{1},k_{1}-\bar{k}_{1}+Q)\sigma^{\mathrm{z}}\bigg],\\
\Pi^{\mathrm{NLO}}(\bar{\omega}_{1}\bar{k}_{1},\bar{\omega}_{2}\bar{k}_{2})&=&
-\beta^{-6}\frac{1}{(\sqrt{N})^{4}}\sum_{\omega_{1}k_{1}...\omega_{4}k_{4}}\sum_{\omega_{3}^{\prime}k_{3}{\prime}}
\sum_{\omega_{4}^{\prime}k_{4}{\prime}}U^{4}\nonumber
\\&&\times G(\omega_{3}-\omega_{3}^{\prime},k_{3}-k_{3}^{\prime}-Q;\omega_{4}^{\prime}-\omega_{4},k_{4}^{\prime}-k_{4}-Q)\nonumber
\\&&\times\mathrm{tr}\bigg[S(\omega_{1}k_{1},\omega_{4}k_{4})\sigma^{\mathrm{z}}
S(\omega_{4}^{\prime},k_{4}^{\prime};\omega_{2}+\bar{\omega}_{2},k_{2}+\bar{k}_{2}-Q)\nonumber
\\&&\times\sigma^{\mathrm{z}}S(\omega_{2}k_{2},\omega_{3}k_{3})\sigma^{\mathrm{z}}\nonumber
\\&&\times S(\omega_{3}^{\prime},k_{3}^{\prime};
\omega_{1}-\bar{\omega}_{1},k_{1}-\bar{k}_{1}+Q)\sigma^{\mathrm{z}}\bigg],\label{SEBNLOM}
\end{eqnarray}
where $Q=(\pi,\pi)^{\mathrm{T}}$ (in units of the inverse lattice
spacing) is the nesting vector in two dimensions.

In order to simplify our calculations, we assume that the
anti-ferromagnetic order parameter $B$ in Eq.~(\ref{Gamma}) is
constant. Then the inverse of the fermion propagator can be
expressed as the following formalism:
\begin{equation}
S^{-1}_{\alpha_{1}\alpha_{2}}(\omega_{1}k_{1},\omega_{2}k_{2})=\beta
C(\omega_{1},k_{1})\delta_{\omega_{1}\omega_{2}}\delta_{k_{1}k_{2}}\delta_{\alpha_{1}\alpha_{2}}
\!\!\!+\!\!\beta
D(\omega_{1},k_{1})\delta_{\omega_{1}\omega_{2}}\delta_{k_{1},k_{2}+Q}\sigma^{\mathrm{z}}_{\alpha_{1}\alpha_{2}}.\label{}
\end{equation}
It then follows that
\begin{equation}
S_{\alpha_{1}\alpha_{2}}(\omega_{1}k_{1},\omega_{2}k_{2})=\beta
\bar{C}(\omega_{1},k_{1})\delta_{\omega_{1}\omega_{2}}\delta_{k_{1}k_{2}}\delta_{\alpha_{1}\alpha_{2}}
\!\!\!+\!\!\beta
\bar{D}(\omega_{1},k_{1})\delta_{\omega_{1}\omega_{2}}\delta_{k_{1},k_{2}+Q}\sigma^{\mathrm{z}}_{\alpha_{1}\alpha_{2}},\label{S12}
\end{equation}
with
\begin{eqnarray}
\bar{C}(\omega_{1},k_{1})&=&\frac{C(\omega_{1},k_{1}+Q)}{C(\omega_{1},k_{1})C(\omega_{1},k_{1}+Q)-D(\omega_{1},k_{1})D(\omega_{1},k_{1}+Q)},
\\\bar{D}(\omega_{1},k_{1})&=&\frac{-D(\omega_{1},k_{1})}{C(\omega_{1},k_{1})C(\omega_{1},k_{1}+Q)-D(\omega_{1},k_{1})D(\omega_{1},k_{1}+Q)}.\label{}
\end{eqnarray}
Similarly, for the boson field one has
\begin{eqnarray}
G^{-1}(\omega_{1}k_{1},\omega_{2}k_{2})&=&\beta
A(\omega_{1},k_{1})\delta_{\omega_{1}\omega_{2}}\delta_{k_{1}k_{2}},
\\G(\omega_{1}k_{1},\omega_{2}k_{2})&=&\beta
A^{-1}(\omega_{1},k_{1})\delta_{\omega_{1}\omega_{2}}\delta_{k_{1}k_{2}}.\label{G12}
\end{eqnarray}
Substituting Eqs.~(\ref{S12}) and (\ref{G12}) into
Eqs.~(\ref{SEFLOM})---(\ref{SEBNLOM}), we obtain
\begin{eqnarray}
\Sigma^{\mathrm{LO}}_{\alpha_{1},\alpha_{2}}(\bar{\omega}_{1}\bar{k}_{1},\bar{\omega}_{2}\bar{k}_{2})&=&
\Sigma^{\mathrm{LO}}_{+}(\bar{\omega}_{1},\bar{k}_{1})\delta_{\bar{\omega}_{1}\bar{\omega}_{2}}
\delta_{\bar{k}_{1}\bar{k}_{2}}\delta_{\alpha_{1}\alpha_{2}}\nonumber
\\&&+\Sigma^{\mathrm{LO}}_{-}(\bar{\omega}_{1},\bar{k}_{1})\delta_{\bar{\omega}_{1}\bar{\omega}_{2}}
\delta_{\bar{k}_{1},\bar{k}_{2}+Q}\sigma^{\mathrm{z}}_{\alpha_{1}\alpha_{2}},\label{Eq49}\\
\Sigma^{\mathrm{NLO}}_{\alpha_{1},\alpha_{2}}(\bar{\omega}_{1}\bar{k}_{1},\bar{\omega}_{2}\bar{k}_{2})&=&
\Sigma^{\mathrm{NLO}}_{+}(\bar{\omega}_{1},\bar{k}_{1})\delta_{\bar{\omega}_{1}\bar{\omega}_{2}}
\delta_{\bar{k}_{1}\bar{k}_{2}}\delta_{\alpha_{1}\alpha_{2}}\nonumber
\\&&+\Sigma^{\mathrm{NLO}}_{-}(\bar{\omega}_{1},\bar{k}_{1})\delta_{\bar{\omega}_{1}\bar{\omega}_{2}}
\delta_{\bar{k}_{1},\bar{k}_{2}+Q}\sigma^{\mathrm{z}}_{\alpha_{1}\alpha_{2}},\label{Eq50}\\
\Pi^{\mathrm{LO}}(\bar{\omega}_{1}\bar{k}_{1},\bar{\omega}_{2}\bar{k}_{2})&=&\Pi^{\mathrm{LO}}(\bar{\omega}_{1},\bar{k}_{1})\delta_{\bar{\omega}_{1}\bar{\omega}_{2}}
\delta_{\bar{k}_{1}\bar{k}_{2}},\\
\Pi^{\mathrm{NLO}}(\bar{\omega}_{1}\bar{k}_{1},\bar{\omega}_{2}\bar{k}_{2})&=&\Pi^{\mathrm{NLO}}(\bar{\omega}_{1},\bar{k}_{1})\delta_{\bar{\omega}_{1}\bar{\omega}_{2}}
\delta_{\bar{k}_{1}\bar{k}_{2}},\label{Eq52}
\end{eqnarray}
with
\begin{eqnarray}
\Sigma^{\mathrm{LO}}_{+}(\bar{\omega}_{1},\bar{k}_{1})&=&U^{2}\frac{1}{(\sqrt{N})^{2}}
\sum_{\omega_{1}k_{1}}A^{-1}(\omega_{1},k_{1})\nonumber
\\&&\times\bar{C}(\bar{\omega}_{1}-\omega_{1},\bar{k}_{1}-k_{1}-Q)
e^{-i(\bar{\omega}_{1}-\omega_{1})0^{+}},\label{Eq53}
\\\Sigma^{\mathrm{LO}}_{-}(\bar{\omega}_{1},\bar{k}_{1})&=&U^{2}\frac{1}{(\sqrt{N})^{2}}
\sum_{\omega_{1}k_{1}}A^{-1}(\omega_{1},k_{1})\nonumber
\\&&\times\bar{D}(\bar{\omega}_{1}-\omega_{1},\bar{k}_{1}-k_{1}-Q)
e^{-i(\bar{\omega}_{1}-\omega_{1})0^{+}},\label{Eq54}\\
\Sigma^{\mathrm{NLO}}_{+}(\bar{\omega}_{1},\bar{k}_{1})&=&U^{4}\beta^{-1}\frac{1}{(\sqrt{N})^{4}}
\sum_{\omega_{1}k_{1}}\sum_{\omega_{2}k_{2}}A^{-1}(\omega_{1},k_{1})A^{-1}(\omega_{2},k_{2})\nonumber
\\&&\times\bigg[\bar{C}(\bar{\omega}_{1}-\omega_{1},\bar{k}_{1}-k_{1}-Q)
\bar{C}(\bar{\omega}_{1}-\omega_{1}-\omega_{2},\bar{k}_{1}-k_{1}-k_{2})\nonumber
\\&&\times\bar{C}(\bar{\omega}_{1}-\omega_{2},\bar{k}_{1}-k_{2}-Q)
+\bar{C}(\bar{\omega}_{1}-\omega_{1},\bar{k}_{1}-k_{1}-Q)\nonumber
\\&&\times\bar{D}(\bar{\omega}_{1}-\omega_{1}-\omega_{2},\bar{k}_{1}-k_{1}-k_{2})
\bar{D}(\bar{\omega}_{1}-\omega_{2},\bar{k}_{1}-k_{2})\nonumber
\\&&+\bar{D}(\bar{\omega}_{1}-\omega_{1},\bar{k}_{1}-k_{1}-Q)\nonumber
\\&&\times\bar{C}(\bar{\omega}_{1}-\omega_{1}-\omega_{2},\bar{k}_{1}-k_{1}-k_{2}-Q)
\bar{D}(\bar{\omega}_{1}-\omega_{2},\bar{k}_{1}-k_{2})\nonumber
\\&&+\bar{D}(\bar{\omega}_{1}-\omega_{1},\bar{k}_{1}-k_{1}-Q)\nonumber
\\&&\times\bar{D}(\bar{\omega}_{1}-\omega_{1}-\omega_{2},\bar{k}_{1}-k_{1}-k_{2}-Q)
\nonumber
\\&&\times\bar{C}(\bar{\omega}_{1}-\omega_{2},\bar{k}_{1}-k_{2}-Q)\bigg],\label{Eq55}\\
\Sigma^{\mathrm{NLO}}_{-}(\bar{\omega}_{1},\bar{k}_{1})&=&U^{4}\beta^{-1}\frac{1}{(\sqrt{N})^{4}}
\sum_{\omega_{1}k_{1}}\sum_{\omega_{2}k_{2}}A^{-1}(\omega_{1},k_{1})A^{-1}(\omega_{2},k_{2})\nonumber
\\&&\times\bigg[\bar{C}(\bar{\omega}_{1}-\omega_{1},\bar{k}_{1}-k_{1}-Q)
\bar{C}(\bar{\omega}_{1}-\omega_{1}-\omega_{2},\bar{k}_{1}-k_{1}-k_{2})\nonumber
\\&&\times\bar{D}(\bar{\omega}_{1}-\omega_{2},\bar{k}_{1}-k_{2}-Q)
+\bar{C}(\bar{\omega}_{1}-\omega_{1},\bar{k}_{1}-k_{1}-Q)\nonumber
\\&&\times\bar{D}(\bar{\omega}_{1}-\omega_{1}-\omega_{2},\bar{k}_{1}-k_{1}-k_{2})
\bar{C}(\bar{\omega}_{1}-\omega_{2},\bar{k}_{1}-k_{2})\nonumber
\\&&+\bar{D}(\bar{\omega}_{1}-\omega_{1},\bar{k}_{1}-k_{1}-Q)\nonumber
\\&&\times\bar{C}(\bar{\omega}_{1}-\omega_{1}-\omega_{2},\bar{k}_{1}-k_{1}-k_{2}-Q)
\bar{C}(\bar{\omega}_{1}-\omega_{2},\bar{k}_{1}-k_{2})\nonumber
\\&&+\bar{D}(\bar{\omega}_{1}-\omega_{1},\bar{k}_{1}-k_{1}-Q)\nonumber
\\&&\times\bar{D}(\bar{\omega}_{1}-\omega_{1}-\omega_{2},\bar{k}_{1}-k_{1}-k_{2}-Q)
\nonumber
\\&&\times\bar{D}(\bar{\omega}_{1}-\omega_{2},\bar{k}_{1}-k_{2}-Q)\bigg]\label{Eq56},\label{}
\end{eqnarray}
and
\begin{eqnarray}
\Pi^{\mathrm{LO}}(\bar{\omega}_{1},\bar{k}_{1})&=&-2U^{2}\frac{1}{(\sqrt{N})^{2}}
\sum_{\omega_{1}k_{1}}\bigg[\bar{C}(\omega_{1},k_{1})\bar{C}(\omega_{1}-\bar{\omega}_{1},k_{1}-\bar{k}_{1}+Q)
\nonumber
\\&&+\bar{D}(\omega_{1},k_{1})\bar{D}(\omega_{1}-\bar{\omega}_{1},k_{1}-\bar{k}_{1})\bigg],\label{}
\end{eqnarray}
\begin{eqnarray}
\Pi^{\mathrm{NLO}}(\bar{\omega}_{1},\bar{k}_{1})&=&-2U^{4}\beta^{-1}\frac{1}{(\sqrt{N})^{4}}
\sum_{\omega_{1}k_{1}}\sum_{\omega_{2}k_{2}}\bigg[A^{-1}(\omega_{2}-\omega_{1}+\bar{\omega}_{1},k_{2}-k_{1}+\bar{k}_{1})\nonumber
\\&&\times\bar{C}(\omega_{1},k_{1})\bar{C}(\omega_{2}+\bar{\omega}_{1},k_{2}+\bar{k}_{1}-Q)\bar{C}(\omega_{2},k_{2})
\nonumber
\\&&\times\bar{C}(\omega_{1}-\bar{\omega}_{1},k_{1}-\bar{k}_{1}+Q)\nonumber
\\&&+A^{-1}(\omega_{2}-\omega_{1}+\bar{\omega}_{1},k_{2}-k_{1}+\bar{k}_{1})\bar{C}(\omega_{1},k_{1})\nonumber
\\&&\times\bar{C}(\omega_{2}+\bar{\omega}_{1},k_{2}+\bar{k}_{1}-Q)
\bar{D}(\omega_{2},k_{2})
\bar{D}(\omega_{1}-\bar{\omega}_{1},k_{1}-\bar{k}_{1})\nonumber
\\&&+A^{-1}(\omega_{2}-\omega_{1}+\bar{\omega}_{1},k_{2}-k_{1}+\bar{k}_{1})\bar{D}(\omega_{1},k_{1})\nonumber\\
&&\times\bar{D}(\omega_{2}+\bar{\omega}_{1},k_{2}+\bar{k}_{1})
\bar{C}(\omega_{2},k_{2})
\bar{C}(\omega_{1}-\bar{\omega}_{1},k_{1}-\bar{k}_{1}+Q)\nonumber\\
&&+A^{-1}(\omega_{2}-\omega_{1}+\bar{\omega}_{1},k_{2}-k_{1}+\bar{k}_{1})
\bar{D}(\omega_{1},k_{1})\nonumber\\
&&\times\bar{D}(\omega_{2}+\bar{\omega}_{1},k_{2}+\bar{k}_{1})
\bar{D}(\omega_{2},k_{2})\bar{D}(\omega_{1}-\bar{\omega}_{1},k_{1}-\bar{k}_{1})\nonumber\\
&&
+A^{-1}(\omega_{2}-\omega_{1}+\bar{\omega}_{1},k_{2}-k_{1}+\bar{k}_{1}-Q)\bar{C}(\omega_{1},k_{1})\nonumber\\
&&\times\bar{D}(\omega_{2}+\bar{\omega}_{1},k_{2}+\bar{k}_{1})
\bar{C}(\omega_{2},k_{2})\bar{D}(\omega_{1}-\bar{\omega}_{1},k_{1}-\bar{k}_{1})\nonumber\\
&&+A^{-1}(\omega_{2}-\omega_{1}+\bar{\omega}_{1},k_{2}-k_{1}+\bar{k}_{1}-Q)
\bar{C}(\omega_{1},k_{1})\nonumber\\
&&\times\bar{D}(\omega_{2}+\bar{\omega}_{1},k_{2}+\bar{k}_{1})\bar{D}(\omega_{2},k_{2})\bar{C}(\omega_{1}-\bar{\omega}_{1},k_{1}-\bar{k}_{1}+Q)
\nonumber\\
&&+A^{-1}(\omega_{2}-\omega_{1}+\bar{\omega}_{1},k_{2}-k_{1}+\bar{k}_{1}-Q)\bar{D}(\omega_{1},k_{1})\nonumber\\
&&\times\bar{C}(\omega_{2}+\bar{\omega}_{1},k_{2}+\bar{k}_{1}-Q)
\bar{C}(\omega_{2},k_{2})\bar{D}(\omega_{1}-\bar{\omega}_{1},k_{1}-\bar{k}_{1})\nonumber\\
&&+A^{-1}(\omega_{2}-\omega_{1}+\bar{\omega}_{1},k_{2}-k_{1}+\bar{k}_{1}-Q)
\bar{D}(\omega_{1},k_{1})\nonumber\\
&&\times\bar{C}(\omega_{2}+\bar{\omega}_{1},k_{2}+\bar{k}_{1}-Q)\nonumber\\
&&\times\bar{D}(\omega_{2},k_{2})\bar{C}(\omega_{1}-\bar{\omega}_{1},k_{1}-\bar{k}_{1}+Q)\bigg].\label{Eq58}\label{}
\end{eqnarray}
Furthermore, from Eq.~(\ref{G0S0}) it is easy to obtain
\begin{eqnarray}
G^{-1}_{0}(\omega_{1}k_{1},\omega_{2}k_{2})&=&\beta
U\delta_{\omega_{1}\omega_{2}}\delta_{k_{1}k_{2}},\label{Eq59}\\
S^{-1}_{0;\alpha_{1}\alpha_{2}}(\omega_{1}k_{1},\omega_{2}k_{2})&=&\beta
(-i\omega_{1}+\xi_{k_{1}})\delta_{\omega_{1}\omega_{2}}\delta_{k_{1}k_{2}}\delta_{\alpha_{1}\alpha_{2}},\\
\Sigma_{\mathrm{mean};\alpha_{1},\alpha_{2}}(\omega_{1}k_{1},\omega_{2}k_{2})&=&
\beta U
B\delta_{\omega_{1}\omega_{2}}\delta_{k_{1},k_{2}+Q}\sigma^{\mathrm{z}}_{\alpha_{1}\alpha_{2}},\label{Eq61}
\end{eqnarray}
with
\begin{equation}
\xi_{k}=-2t(\cos k_{x}+\cos k_{y})-\mu,\label{}
\end{equation}
where the momenta $k_{x}$ and $k_{y}$ are in units of the inverse
lattice spacing. Substituting Eqs.~(\ref{Eq49})---(\ref{Eq52}) and
Eqs.~(\ref{Eq59})---(\ref{Eq61}) into the self-consistent
equations~(\ref{FermiEq}) and~(\ref{BoseEq}), we finally get
\begin{eqnarray}
A(\omega,k)&=&U-\beta^{-1}\Pi^{\mathrm{LO}}(\omega,k)-\beta^{-1}\Pi^{\mathrm{NLO}}(\omega,k),\label{EqA}\\
C(\omega,k)&=&-i\omega+\xi_{k}-\beta^{-1}\Sigma^{\mathrm{LO}}_{+}(\omega,k)-\beta^{-1}\Sigma^{\mathrm{NLO}}_{+}(\omega,k),\\
D(\omega,k)&=&-UB-\beta^{-1}\Sigma^{\mathrm{LO}}_{-}(\omega,k)-\beta^{-1}\Sigma^{\mathrm{NLO}}_{-}(\omega,k).\label{EqD}\label{}
\end{eqnarray}

\section{Numerical Results}
\label{Numericalsection}

Besides equations~(\ref{EqA})---(\ref{EqD}), we still need another
two equations to perform the numerical calculations. One is the
self-consistent equation for the antiferromagnetic order parameter
$B$, which can be obtained from the first equation in
Eqs.~(\ref{SelfEq1}), as given by
\begin{equation}
B=-2\beta^{-1}\frac{1}{(\sqrt{N})^{2}} \sum_{\omega
k}\bar{D}(\omega,k);\label{Beq}
\end{equation}
the other one is the equation for the electron density (number per
lattice site):
\begin{equation}
n=-2\beta^{-1}\frac{1}{(\sqrt{N})^{2}} \sum_{\omega
k}\bar{C}(\omega,k)e^{i\omega 0^{+}}.\label{Eqn}
\end{equation}
We employ a $N_{s}\times N_{s}\times N_{\tau}$ lattice, where
$N_{s}$ corresponds to the lattice number in one spatial dimension
and we have two-dimensional lattice number $N=N_{s}\times N_{s}$;
$N_{\tau}$ is the temporal lattice number. The calculations are
performed in the momentum and frequency ($k_{x}, k_{y}, i\omega$)
space, which are equivalent to those in the coordinate space with
spacial periodic boundary conditions and temporal anti-periodic
(periodic) boundary conditions for fermion (boson) fields.

In order to save computing time, we also employ the symmetries of
the lattice to simplify our calculations. Propagators and
self-energies are invariant under the following symmetry operations:
\begin{equation}
k_{x}\leftrightarrow k_{y},\quad k_{x}\rightarrow -k_{x},\quad
k_{y}\rightarrow -k_{y}.\label{}
\end{equation}
Therefore, we only need to compute one-eighth of lattice points in
the Brillouin zone. Furthermore, we also have the time-reversal
symmetry as follows
\begin{eqnarray}
A(-\omega,k)&=&A(\omega,k)^{*},\label{}\\
C(-\omega,k)&=&C(\omega,k)^{*},\\
D(-\omega,k)&=&D(\omega,k)^{*}.\label{}\label{}
\end{eqnarray}
For this reason, it is only necessary to calculate the propagators
when $\omega\geq 0$.

In this work, we want to investigate the importance of higher-order
fluctuations in the 2PI effective action theory. In another word, we
would like to study whether the expansions of the effective action
in order of loops are convergent, when the interaction strength $U$
becomes large. In order to obtain this goal, we perform three
different calculations: the first one is just the mean field
calculation, i.e., neglecting the influence of the fluctuations and
assuming $\Gamma_{2}=0$. The second one is that only the two-loop
contribution (the first diagram in Fig.~\ref{Gamma2Fig}) to
$\Gamma_{2}$ are included. In this calculation,
equations~(\ref{EqA})---(\ref{Eqn}) constitute a closed system and
are solved self-consistently through iterations, but with
$\Pi^{\mathrm{NLO}}=\Sigma^{\mathrm{NLO}}_{+}=\Sigma^{\mathrm{NLO}}_{-}=0$.
As for the third calculation, it should have been solving this set
of equations self-consistently with
$\Pi^{\mathrm{NLO}}=\Sigma^{\mathrm{NLO}}_{+}=\Sigma^{\mathrm{NLO}}_{-}\neq0$,
but such self-consistent calculation need lots of computing time and
is beyond our computing abilities. Therefore, in the third
calculation we substitute the results of the self-consistent
equations which only include the LO contribution to the
fluctuations, i.e., the results obtained from the second
calculation, into equations~(\ref{Eq55}), (\ref{Eq56}), and
(\ref{Eq58}) to obtain the NLO self energies. We should emphasize
that this calculation is not self-consistent, but it is still
reasonable to get some information on the convergence of the theory
from this calculation.

Equations~(\ref{EqA})---(\ref{Eqn}) are iterated successively to
search for self-consistent solutions. In order to avoid oscillations
during the calculations, we also employ feedback in the process of
iterations, i.e., the updated value is chosen to be a weighted
average of the new calculated one and the old one in every
iteration. We use the following criteria to judge whether the
convergence of iterations is achieved:
\begin{eqnarray}
\frac{|A_{\mathrm{new}}(\omega,k)-A_{\mathrm{old}}(\omega,k)|}{|A_{\mathrm{new}}(\omega,k)|}&<&0.001,\label{}\\
\frac{|C_{\mathrm{new}}(\omega,k)-C_{\mathrm{old}}(\omega,k)|}{|C_{\mathrm{new}}(\omega,k)|}&<&0.001,\\
\frac{|D_{\mathrm{new}}(\omega,k)-D_{\mathrm{old}}(\omega,k)|}{|D_{\mathrm{new}}(\omega,k)|}&<&0.001,\label{}\label{}
\end{eqnarray}
for all $\omega$ and $k$, and
\begin{eqnarray}
\frac{|B_{\mathrm{new}}-B_{\mathrm{old}}|}{|B_{\mathrm{new}}|}&<&0.001,\label{}\\
\frac{|\mu_{\mathrm{new}}-\mu_{\mathrm{old}}|}{|\mu_{\mathrm{new}}|}&<&0.001,\label{}\label{}
\end{eqnarray}
where the update of the chemical potential $\mu$ is realized through
a Newton's procedure at a fixed electron density $n$. Furthermore,
we would like to emphasize that there is an exponential factor
appearing in Eqs.~(\ref{Eq53}),~(\ref{Eq54}), and~(\ref{Eqn}). This
exponential factor contains an infinitesimal positive constant
$0^{+}$. In the numerical calculations, this constant can not be
chosen to be very small, because that would affect the convergence
of the frequency summation, since the infinite frequency summation
is cut to be finite in numerical calculations. Therefore, in our
work we choose a relatively small constant, and we also check that
the numerical results are insensitive to the choice of this
constant.

\begin{figure}[!htb]
\includegraphics[scale=0.8]{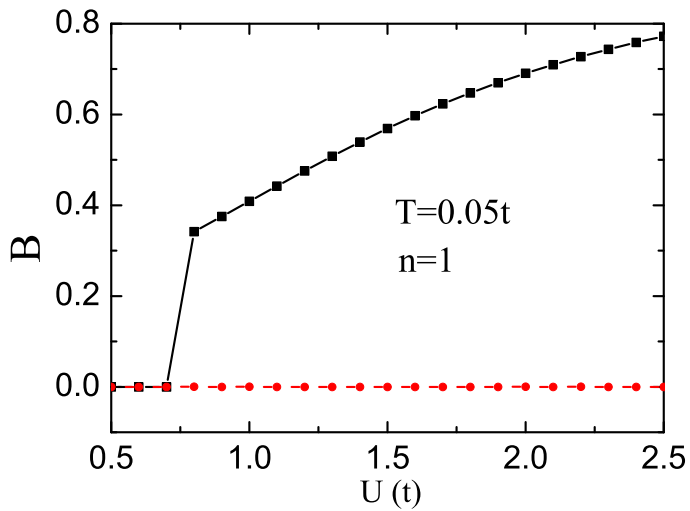}
\caption{(color online). Antiferromagnetic order parameter $B$ as a
function of the Coulomb repulsion $U$ in unit of the hopping $t$.
Here the temperature is chosen to be $T=0.05 t$ and the electron
density $n=1$. The black solid line with squares corresponds to the
mean field calculations and the red dashed line with circles to the
self-consistent calculations including LO fluctuations. For the
self-consistent calculations, the Brillouin zone is discretized into
a $10\times 10$ lattice and the temporal lattice number is chosen to
be $N_{\tau}=64$.}\label{BUFig}
\end{figure}

\begin{figure}[!htb]
\includegraphics[scale=0.8]{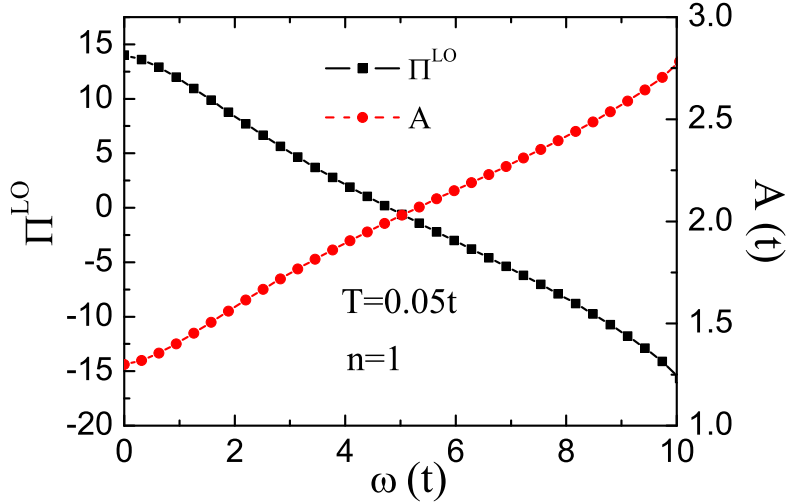}
\caption{(color online). Dependence of the LO boson self-energy
$\Pi^{\mathrm{LO}}(\omega,k_{x}=0,k_{y}=0)$ and the inverse of the
boson propagator $A(\omega,k_{x}=0,k_{y}=0)$ on the Matsubara
frequencies, obtained in the self-consistent calculations with LO
fluctuations included. The Coulomb repulsion is $U=2t$.  These
results are based on a lattice of $10\times 10\times
64$.}\label{PiLOAFig}
\end{figure}

The antiferromagnetic order parameter $B$ as a function of $U$,
obtained in the mean field approximations and the self-consistent
calculations including LO fluctuations, are shown in
Fig.~\ref{BUFig}. In the mean field approximations, the frequency
summation in Eqs.~(\ref{Beq}) and~(\ref{Eqn}) can be performed
analytically. One can see that $B$ develops a nonvanishing value
with the increase of the Coulomb repulsion in the mean field
approximations, which means that a phase transition occurs and a
long-range antiferromagnetic state is formed. However, in the
self-consistent calculations with LO fluctuations included, we do
not find any phase transition at $T=0.05 t$, $n=1$, and $U=0.5\sim
2.5t$. In order to find the reason why there is no phase transition
in the self-consistent calculations, we show the LO boson
self-energy and the inverse of the boson propagator $A$ in
Fig.~\ref{PiLOAFig}. We can observe that $\Pi^{\mathrm{LO}}$ is
positive at low frequencies, which results in that $A$ is decreased
at low frequencies compared to the bare Coulomb repulsion $U$.
Therefore, the effective interaction is reduced by the fluctuations
at low frequencies, which may account for the reason why the phase
transition becomes difficult in the self-consistent calculations
with fluctuations included.

\begin{figure}[!htb]
\includegraphics[scale=0.9]{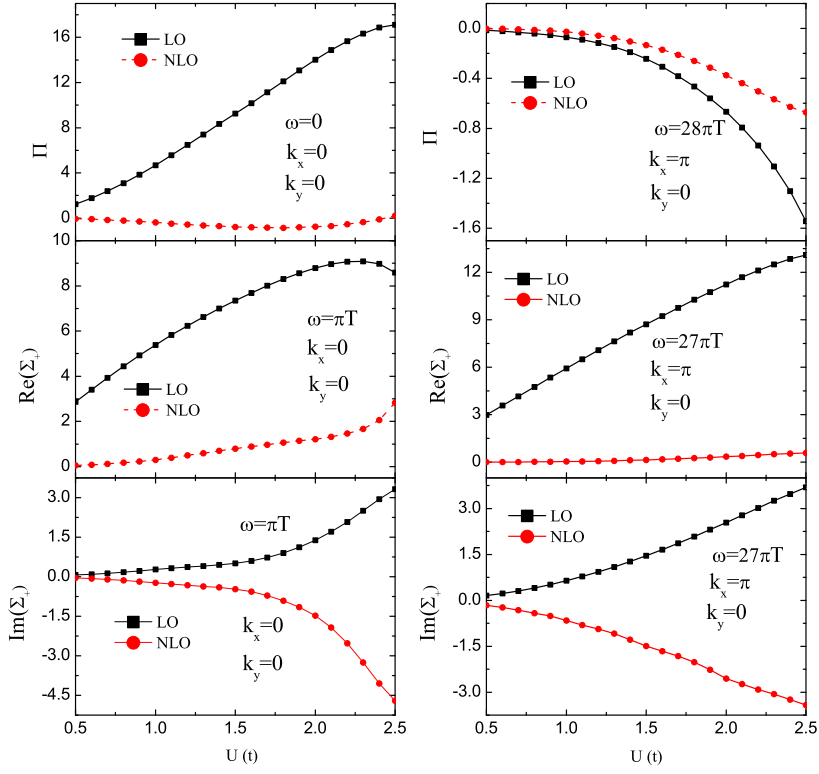}
\caption{(color online). Comparison between the LO self-energies and
the NLO ones. They are depicted as functions of the Coulomb
repulsion $U$. Temperature $T=0.05 t$ and electron density $n=1$ are
chosen here. $k_{x}$ and $k_{y}$ are in unit of the inverse of the
lattice spacing. These results are based on a lattice of $10\times
10\times 64$.}\label{PiSigmaUFig}
\end{figure}

Figure~\ref{PiSigmaUFig} shows the LO and NLO self-energies as
functions of the on-site repulsion $U$. The boson self-energies
$\Pi$ are real and are presented in the two top panels. In the
symmetrical phase where $B=0$, $\Sigma_{-}$ defined in
Eqs.~(\ref{Eq49}) and~(\ref{Eq50}) is vanishing. Therefore, we only
show $\Sigma_{+}$ here,  whose real and imaginary parts are depicted
in the two medium panels and the two bottom ones of
Fig.~\ref{PiSigmaUFig}, respectively. From this figure, one can
observe that the NLO contributions to $\Pi$ and the real part of
$\Sigma_{+}$ are less than the LO ones, but as the $U$ is larger
than about $2t$, the NLO contributions can not be neglected as the
top-right and medium-left panels show. However, for the imaginary
part of the $\Sigma_{+}$, we can see that the NLO contributions are
completely comparable with the LO ones, even when the $U$ is less
than $2t$.

\begin{figure}[!htb]
\includegraphics[scale=1.2]{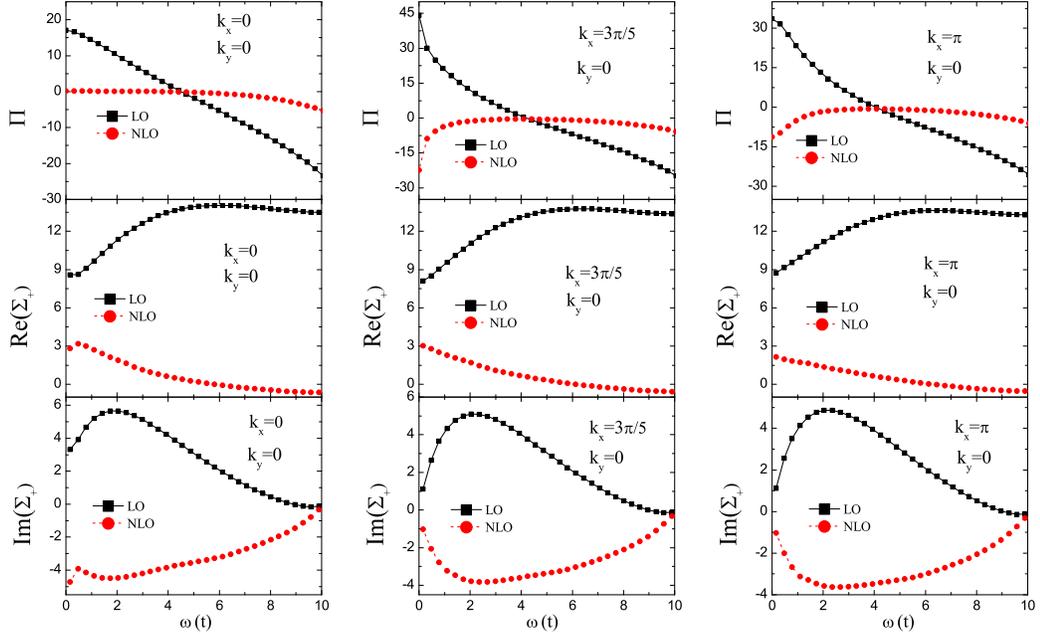}
\caption{(color online). LO and NLO self-energies as functions of
the frequencies. Three columns correspond to three different values
of $k_{x}$. In the same way, we choose the temperature $T=0.05 t$,
electron density $n=1$, and the Coulomb repulsion $U=2.5t$. The
lattice is $10\times 10\times 64$.}\label{PiSigmaOmegaFig}
\end{figure}

In order to make the comparison much clearer, we show the spectrum
of the self-energies in Fig.~\ref{PiSigmaOmegaFig}. Here we choose
$U=2.5t$. For the boson self-energies, we find that the NLO
contributions are significant and should not be neglected,
especially at low frequencies as the top-middle and top-right panels
show. In the same way, the same conclusions are also appropriate for
the real part of the $\Sigma_{+}$. Furthermore, we also find that
the real part of the NLO $\Sigma_{+}$ approaches zero at high
frequencies. For the imaginary part of the $\Sigma_{+}$, we confirm
that the NLO contributions are comparable with the LO ones, which is
clear at the whole region of the frequencies as the bottom three
panels show.

\begin{figure}[!htb]
\includegraphics[scale=1.2]{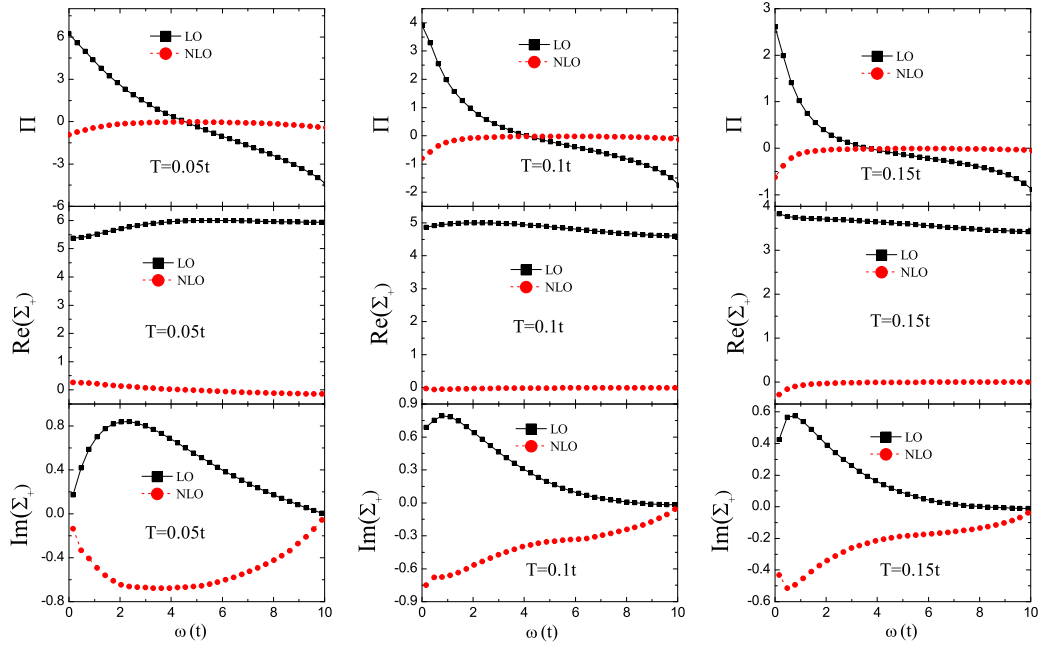}
\caption{(color online). LO and NLO self-energies as functions of
the frequencies at three different values of the temperature. We
choose $k_{x}=3\pi/5$, $k_{y}=0$, $U=1t$, and $n=1$ in all these
calculations. These results are based on a lattice of $10\times
10\times 64$.}\label{PiSigmaTFig}
\end{figure}

Figure~\ref{PiSigmaTFig} depicts the self-energies at several values
of the temperature. Here we choose the Coulomb repulsion $U=t$. We
have demonstrated above that at this value of $U$, the NLO
contributions to the boson self-energy $\Pi$ and the real part of
the fermion self-energy $\Sigma_{+}$ are quite smaller than the LO
ones, which are confirmed at other values of temperature as the
first two rows of Fig.~\ref{PiSigmaTFig} show. In the same way, at
other values of temperature, we also find that the NLO contributions
to the imaginary part of the fermion self-energy are comparable to
those of LO.

\section{Summary and Conclusions}
\label{concSect}

In this work, we have employed the 2PI effective action theories to
study the strongly fluctuating electron systems, under the formalism
of the two-dimensional Hubbard model. We first bosonize the original
classic action of the Hubbard model, then obtain the corresponding
quantum 2PI effective action. In our actual calculations, the 2PI
effective action is expanded to three loops. Therefore, LO and NLO
quantum fluctuations are included in our approaches. We also obtain
the LO and NLO contributions to the fermion and boson self-energies.
They are expressed in the momentum-frequency space in a form which
is very appropriate for numerical calculations.

We also perform numerical calculations on a lattice. Our numerical
results indicate that due to the decrease of the effective Coulomb
repulsion at low frequencies when the quantum fluctuations are
included, the state with an antiferromagnetic long-range order is
more difficult to formed. We also compare the LO and NLO
contributions to the fermion and boson self-energies. We find that
the NLO contributions to the boson self-energy and the real part of
the fermion self-energy are less than the LO ones, but as the
Coulomb repulsion energy $U$ is larger than about $2t$, the NLO
contributions can not be neglected. However, for the imaginary part
of the fermion self-energy, it is found that the NLO contributions
are comparable with the LO ones, even when the $U$ is less than
$2t$. However, their signs are opposite and their sum almost
approaches zero.

Based on our calculations, we conclude that the 2PI effective action
formalism, which is popularly employed in the particle physics and
field theories, is also an appropriate approach to describe the
strongly correlated electron systems. Higher-order quantum
fluctuations are easily included in this approach. We must point out
that one should be very careful about results obtained from the
mean-field calculations of the strongly correlated electron systems,
or even those including the LO quantum fluctuations. Because
higher-order quantum fluctuations in the systems also play an
important role, which can not be neglected.

\section*{Acknowledgements}

This work was supported by the National Natural Science Foundation
of China under Contracts No. 11005138.





\bibliographystyle{model1a-num-names}
\bibliography{<your-bib-database>}



\end{document}